\documentclass{PoS}

\usepackage{latexsym}
\usepackage{epstopdf}
\usepackage{epsfig}
\usepackage{multicol}
\usepackage{pstricks,pst-grad}
\usepackage{cite}

\usepackage{amsmath}
\usepackage{amsfonts}
\usepackage{amssymb}
\usepackage{phonetic}
\usepackage{rotating}

\newcommand{\be}{ \begin{equation} }
\newcommand{\ee}{\end{equation}}
\newcommand{\bea}{\begin{eqnarray}}
\newcommand{\eea}{\end{eqnarray}}
\newcommand{\bit}{\begin{itemize}}
\newcommand{\eit}{\end{itemize}}
\newcommand{\bmp}[1]{\begin{minipage}{#1cm}}
\newcommand{\emp}{\end{minipage}}

\newcommand{\dd}{\mbox{\hausad}}
\newcommand{\Tr}{\mbox{Tr}}
\newcommand{\Id}{1\!\!1}
\newcommand{\half}{\frac{1}{2}}

\newcommand{\bra}{\langle}
\newcommand{\ket}{\rangle}

\title{Adaptive gauge cooling for complex Langevin dynamics}
\ShortTitle{Adaptive gauge cooling for complex Langevin dynamics}

\author{\speaker{Lorenzo Bongiovanni} \\
Department of Physics, College of Science, Swansea University, Swansea, United Kingdom\\
E-mail: \email{pylb@pyserver.swan.ac.uk}}
\author{Gert Aarts\\
Department of Physics, College of Science, Swansea University, Swansea, United Kingdom\\
E-mail: \email{G.Aarts@swansea.ac.uk}}
\author{Erhard Seiler\\
Max-Planck-Institut f\"ur Physik, M\"unchen, Germany\\
E-mail: \email{ehs@mpp.mpg.de}}
\author{D\'enes Sexty\\
Institut f\"ur Theoretische Physik, Universit\"at Heidelberg, Heidelberg, Germany\\
E-mail: \email{D.Sexty@ThPhys.Uni-Heidelberg.DE}}
\author{Ion-Olimpiu Stamatescu\\
Institut f\"ur Theoretische Physik, Universit\"at Heidelberg, Heidelberg, Germany\\
E-mail: \email{stamates@ThPhys.Uni-Heidelberg.DE}}

\abstract{
In the case of nonabelian gauge theories with a complex weight, a controlled exploration of the complexified configuration space during a complex Langevin process requires the use of SL($N, \mathbb{C}$) gauge cooling, in order to minimize the distance from SU($N$) . Here we show that adaptive gauge cooling can lead to an efficient implementation of this idea. First results for SU(3) Yang-Mills theory in the presence of a nonzero $\theta$-term are presented as well.
}

\FullConference{31st International Symposium on Lattice Field Theory
LATTICE 2013\\
July 29 -- August 3, 2013\\
Mainz, Germany}

\begin{document}

\section{Introduction}

The lattice discretisation provides an important tool to study QCD nonperturbatively. However standard methods, based on importance sampling, require a real Boltzmann weight to be used as a probability distribution. Therefore they cannot be used to investigate many problems of great interest where the weight is actually complex, such as in the case of the QCD phase diagram at nonzero baryon chemical potential and QCD in the presence of a $\theta$-term.
 The issue of a complex weight goes under the name of the {\em sign problem} \cite{Aarts:2013bla}. 
 Complex Langevin (CL) dynamics \cite{Parisi:1984cs} offers, in principle, a way to carry out simulations without requiring the weight to be real, since importance sampling is not used.  Although real Langevin dynamics is proven to converge to the desired distribution, there is not (yet) such a general proof for CL.
 Recently, a formal proof of convergence was found that depends, however, on sufficiently tight localisation of the probability distribution in the 
complex configuration space \cite{Aarts:2009uq}.
In the context of gauge theories this implies that the degrees of freedom should not venture out too widely into the enlarged SL$(N,\mathbb{C})$ group.
It is then clear that controlling the dynamics in such a way that it remains localised in the complexified field space  becomes a crucial point to ensure its convergence to the right result.  In particular, for nonabelian gauge theories gauge cooling (GC) is a method that makes progress possible
\cite{Seiler:2012wz,Aarts:2013uxa,Sexty:2013ica}.

\section{Gauge cooling}

In nonabelian gauge theories, complex Langevin dynamics naturally enlarges the gauge group of the theory from SU$(N)$ to SL$(N,\mathbb{C})$, by complexifying its parameters. While the determinant of its elements remains unity, unitarity no longer holds  \cite{Aarts:2008rr},
\be
U \in  \mbox{SL}(N,\mathbb{C}):  \quad\quad\quad\quad  UU^\dag \neq \Id \quad\quad\quad\quad  \dfrac{1}{N}\Tr\, UU^\dag \geq 1.
\ee
Gauge invariance is of course still present: a transformation at site $x$
\be
U_{x,\mu}  \rightarrow  \Omega_x U_{x,\mu}
\quad\quad\quad\quad
U_{x-\hat\mu,\mu}  \rightarrow  U_{x-\hat\mu,\mu}\Omega^{-1}_x,
\ee
where
\be
 \Omega_x =e^{i\omega_{ax}\lambda_a} \in \mbox{SL}(N,\mathbb{C}),
 \quad\quad\quad\quad \omega_{ax} \in \mathbb{C},
\ee
leaves the action invariant. Here $\lambda_a$ are the Gell-Mann matrices  ($a=1,\ldots,N^2-1$, sum over $a$ understood).
However, the gauge orbit, described by $\Omega_x$, now takes place in SL$(N,\mathbb{C})$. Moreover, it changes the unitarity norm (UN) 
$\Tr\, UU^\dag$ (note that $\Tr\, UU^{-1}=N$ is preserved).
The idea of gauge cooling \cite{Seiler:2012wz} is to use this noninvariance of the UN under gauge transformation to move all the links, in a configuration, along the gauge orbit up to the point where the UN is minimal.
One can obtain this by choosing, as parameter of the gauge transformation, the gradient of the UN itself along the gauge orbit. A gauge transformation at site $x$ then takes the form
\be
\label{eq:24}
\Omega_x = e^{-\epsilon \alpha_{\rm gf}f_{ax} \lambda_a}, \quad\quad\quad f_{ax} = 2 \Tr\left[ \lambda_a \left(U_{x,\mu}U_{x,\mu}^\dag -U_{x-\hat\mu,\mu}^\dag U_{x-\hat\mu,\mu} \right)\right],
\ee
where  in the latter expression the sum over all directions is taken.
In the exponent $\epsilon$ is a finite  parameter representing the order of magnitude of the stepsize used in the Langevin process, while
$\alpha_{\rm gf}$  is a parameter which can still be chosen to optimise the cooling.
It is important to stress that since the gauge cooling process is separate from the CL evolution, it is not equivalent to gauge fixing and no Fadeev-Popov determinant is needed.

\begin{figure}[t] 
\centering 
\includegraphics[width=0.4\textwidth]{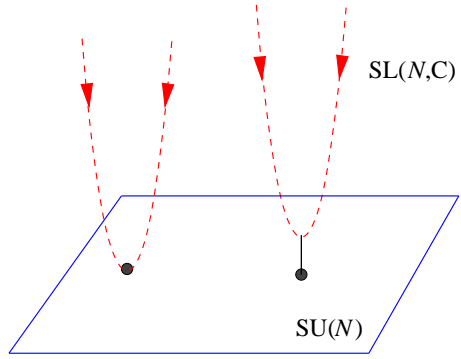} 
\caption{Gauge cooling in SL$(N,\mathbb{C})$ brings the link as close as possible to SU$(N)$. The orbit on the left is equivalent to a SU$(N)$ configuration, while the one on the right is not. } 
\label{fig:SUN orbit} 
\end{figure}

\section{Analytical solution of gauge cooling in a one-link model}

It is instructive to derive the analytical solution of the GC equation in a one-link SU($N$) model \cite{Aarts:2013uxa}.
Let us define the distance from the unitary manifold as
\be\label{d}
\dd = \dfrac{1}{N}\Tr \left( UU^\dag-\Id \right) \geq 0.
\ee
We apply an infinitesimal GC transformation to the link $U \rightarrow U' = \Omega U \Omega^{-1}$, where $\Omega$ is given in Eq.\ (\ref{eq:24}) (since there is only one link, the indices $x,\mu$ can be dropped). 
The change in the distance $\dd$, at leading order in $\epsilon$, reads
\be
\label{eqgc}
\dfrac{\dd'-\dd}{\epsilon} \rightarrow   \dot{\dd}= - \dfrac{\alpha_{\rm gf} f^2_a}{N} 
= - \dfrac{16 \alpha_{\rm gf}}{N} \Tr \left( UU^\dag \left[U,U^\dag \right] \right).
\ee
This expression is correct for $U\in$ SL($N,\mathbb{C}$). We now continue with $N=2$, considering both the case that $U$ is gauge-equivalent to a matrix in SU(2) and the case that it is not.
 In the first example one expects GC to bring the distance $\dd$ back to 0, while in the second case the distance should remain finite. This is illustrated in
  Fig.\ \ref{fig:SUN orbit}.

For $N=2$, the GC equation (\ref{eqgc}) can be expressed in terms of the two invariants $c=\half \Tr\, U$ and $c^*=\half \Tr\, U^\dag$. It reads
\be
\dot{\dd} = -8 \alpha_{\rm gf} \left(\dd^2+2(1-|c|^2)\dd+c^2+c^{*2}-2|c|^2\right).
\ee
Consider now the case that $c=c^*$ (and $c\neq 1$; $c=1$ corresponds to the identity matrix). In that case, $U$ is gauge-equivalent to an element of SU(2) and the GC equation simplifies to
\be
\dot{\dd} = -8 \alpha_{\rm gf} \left(\dd+2(1-c^2)\right)\dd.
\ee
This equation indeed has a  {\em unique} fixed point at  $\dd=0$, which is reached exponentially fast,
\be
 \dd(t)\sim 2 (1-c^2) e^{-16\alpha_{gf}(1-c^2)t} \rightarrow 0.
 \ee
On the other hand, if $c \neq c^*$, $U$ cannot be gauge-equivalent to a SU(2) matrix. The stationary point is 
\be
\dd(t \rightarrow \infty ) =  |c|^2 -1 + \sqrt{1-c^2-c^{*2}+|c|^4} > 0,
\ee
and the minimum distance is larger than 0, as expected.

\begin{figure}[t]
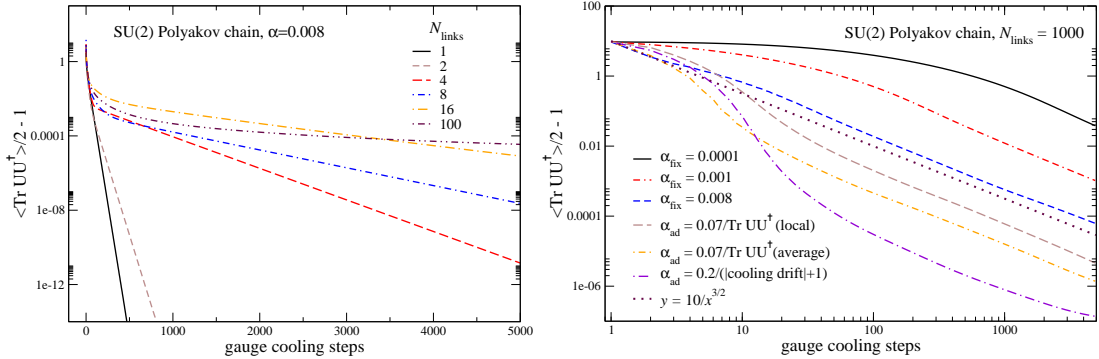
 
\begin{center}
\includegraphics[width=0.465\textwidth]{Nlink-v2.eps} 
\includegraphics[width=0.49\textwidth]{alpha_gf-v6.eps} 
\end{center}
\caption{SU(2) Polyakov chain. Left: Effect of GC on Polyakov chains with a different number of links.
Right: Effect of different implementations of adaptive GC on a Polyakov chain of length $N_\ell=1000$, plotted on a log-log scale.} 
\label{fig:Nlink} 
\end{figure}

\section{Adaptive gauge cooling in the SU(2) Polyakov chain model}

When the model is more complicated than above, analytical solutions are typically not available and a numerical computation is needed. 
The model we consider is an SU(2) Polyakov chain  \cite{Seiler:2012wz,Aarts:2013uxa}, with the action
\be
\label{eq:s}
 S= - \dfrac{\beta}{2}\Tr \left( U_1U_2...U_{N_\ell}\right), \quad\quad\quad \beta \in \mathbb{C}.
\ee
This model can of course be reduced, by an appropriate gauge transformation, to the one-link model $S= - \frac{\beta}{2}\Tr \left( U\right)$ which is exactly solvable. However from a numerical point of view, the action (\ref{eq:s}) has as many degrees of freedom as the number of group parameters times the number of links. In that sense it is a very useful toy model to study the effect of gauge cooling in anticipation of proper four-dimensional gauge theories.

In this exercise, we start with a gauge-transformed unitary chain, such that the initial distance from  SU(2) is nonzero. Under cooling, the distance should then be reduced to zero. In Fig.\ \ref{fig:Nlink} (left) we show how this happens as the number of links $N_\ell$ in the chain is increased.
For $N_\ell=1$ one observes the exponential evolution to the stationary solution as proved anaytically  above. On the other hand, for  
$N_\ell > 1$ the process of reaching  the stationary point is visibly slowed down.

As the number of degrees of freedom increases one wants to find a way for GC to evolve faster towards the stationary fixed point. A possible solution is to use {\em adaptive gauge cooling} \cite{Aarts:2013uxa}, with
\be
\alpha_{\rm ad} = \dfrac{\alpha}{D(U,U^\dag)}.
\ee
Here $D(U,U^\dag)$ is a scalar function that can be adapted to the model under investigation. In Fig.~\ref{fig:Nlink} (right) we show results for several different choices of $D(U,U^\dag)$:
$D_0(U,U^\dag) = 1$ (not adaptive);
$D_1(U,U^\dag) = \Tr\, UU^\dag$;
$D_2(U,U^\dag)=\langle  f_{ax}(U,U^\dag) \rangle_{a,x}$   \cite{Aarts:2013uxa}.
Plotting the results on a log-log scale, one can observe an asymptotic powerlike decay of $\dd(t) \sim t^{-3/2}$. 
As one can appreciate from the plot, with $D_2(U,U^\dag)$  the fixed-point solution is approached several orders of magnitude faster when compared to the nonadaptive one.

\begin{figure}[t]
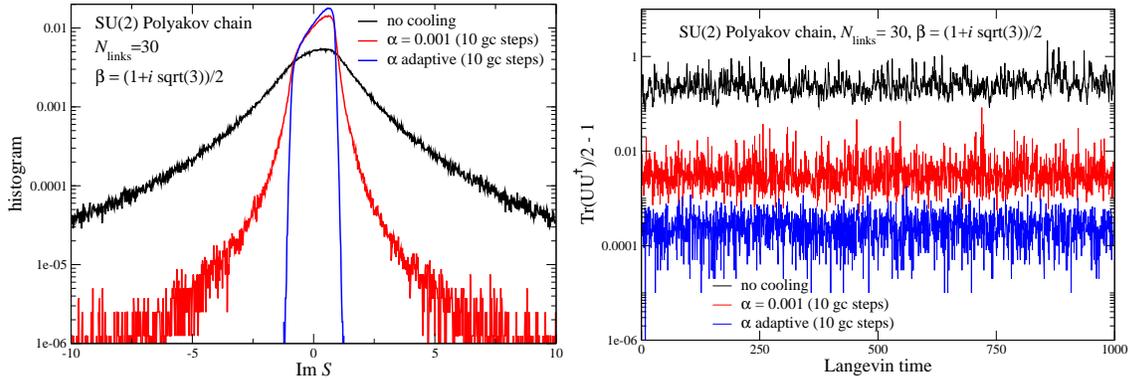
 
\begin{center} 
\includegraphics[width=0.49\textwidth]{ImS_Bin10000.eps} 
\includegraphics[width=0.49\textwidth]{hist-TrUU.eps} 
\end{center}
\caption{SU(2) Polyakov chain. Left: Distribution of the imaginary part of the action for different types of GC. Right: Langevin evolution of the distance $\dd$ for different types of GC.} 
\label{fig:Skirt} 
\end{figure} 

Up to now we have not included actual Langevin dynamics in the evolution. We continue by alternating Langevin updates with cooling steps, in order to carry out a complete simulation. Note that since $\beta$ is chosen to be complex, the action (\ref{eq:s}) is complex-valued and complex Langevin dynamics is required.
However, since the model is equivalent to an exactly solvable one, a comparison between our numerical results and the exact ones is possible.

\begin{figure}[t]
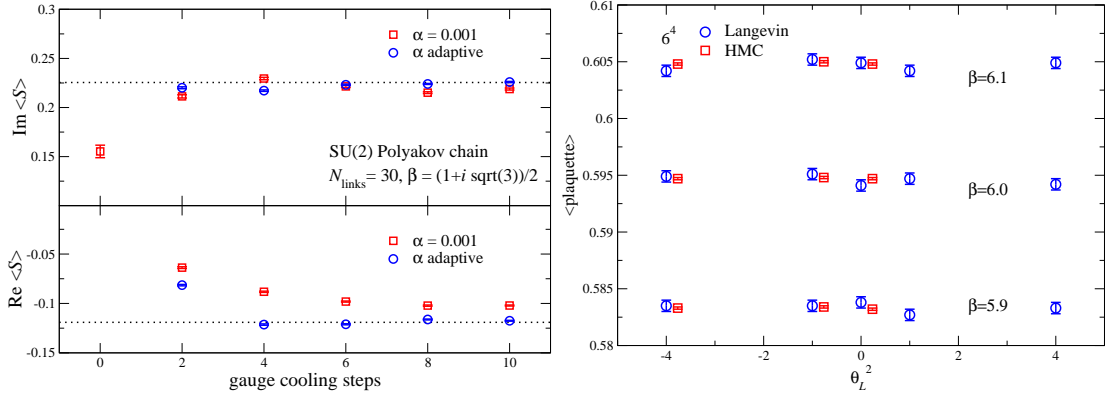
 
\begin{center}
\includegraphics[width=0.48\textwidth]{S-vsgcsteps-v4.eps} 
\includegraphics[width=0.48\textwidth]{Plaq_vs_thetasqr2-v2.eps} 
\end{center}
\caption{Left: Convergence of CL to the correct result after the application of fixed and adaptive GC in the SU(2) Polyakov chain.
Right: expectation values of the plaquette in SU(3) Yang-Mills theory in the presence of a $\theta$-term, for real and imaginary $\theta_L$, using CL and HMC (imaginary $\theta_L$ only), on a $6^4$ lattice at three $\beta$ values. The HMC data has been shifted horizontally for clarity.
}
\label{fig:ExactResult} 
\end{figure}

In Fig.\ \ref{fig:Skirt} (left), we show how GC in general, and the adaptive ones in particular, can constrain the distribution of observables, such as the action, in the complex plane. This is indeed a fundamental  property necessary for correct convergence of CL \cite{Aarts:2009uq}.
The effect on the distance $\dd$ is shown in Fig.\ \ref{fig:Skirt} (right): we observe that during the CL dynamics, adaptive GC can keep the degrees of freedom several orders of magnitude closer to SU(2) than without cooling. We note that since the action is complex, the distance cannot be equal to 0. 
Finally, in Fig.\ \ref{fig:ExactResult} (left) we show how the results for observables, in this case the expectation value of the action, depends on the cooling implementation and the number of cooling steps. 
We observe a clearly convergence to the expected results, indicated with the dotted lines, and the usefulness of adaptive GC.
We also note that initially CL converged to the wrong result. 

An obvious question is how much cooling is required, especially in the case when the exact result is not known. Here we point out that it is not possible  to cool too much, i.e.\ cooling will always bring the configuration closer to SU($N$), until a minimal distance is reached. 
We conclude that in gauge theories  GC is an essential method for the convergence of CL to the correct result and that adaptive GC can help this convergence to be reached quicker.

\begin{figure}[t]
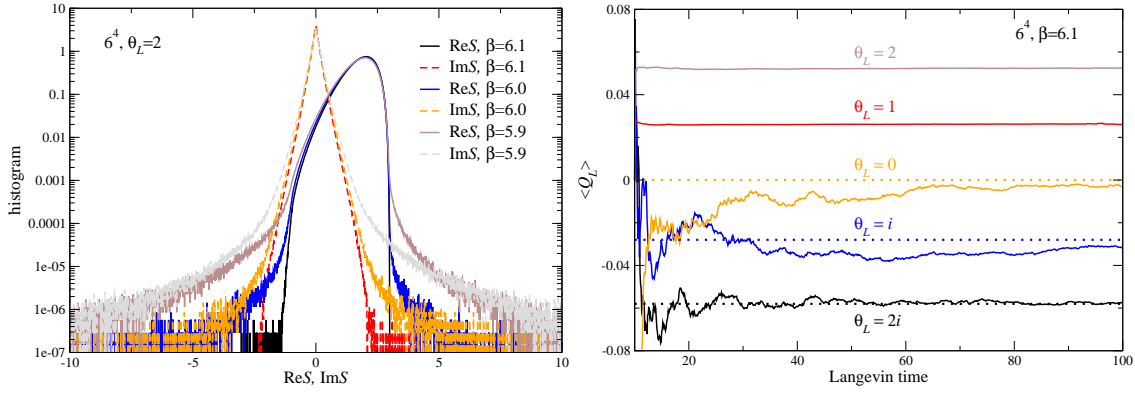

\begin{center}
\includegraphics[width=0.495\textwidth]{skirtS_various_b.eps} 
\includegraphics[width=0.49\textwidth]{Qtop_vs_theta_plusMC-v2.eps} 
\end{center}
\caption{SU(3) Yang-Mills theory in the presence of a $\theta$-term. 
Left: Distribution of the action  for $\theta_L=2$ and  $\beta=5.9, 6, 6.1$.
Right: Running averages of the imaginary (real) part of the topological charge, for real (imaginary) $\theta_L$, using CL, for $\beta=6.1$. The dotted lines for imaginary $\theta_L$ indicate the HMC result. 
 } 
\label{fig:plaq} 
\end{figure}

\section{Preliminary results for 4d SU(3) Yang-Mills theory with a $\theta$-term}

We now apply the method to study a case of physical interest: 
SU(3) Yang-Mills theory in the presence of a $\theta$-term, with the action $S=S_{\rm YM}-i\theta Q$, where $Q$ is the topological charge. To formulate this on the lattice, we follow Refs.\ \cite{Vicari:2008jw,D'Elia:2012vv} and take
\be
\label{eq:st}
  S = S_W  -i\theta_L\sum_x q_L(x),
  \quad\quad\quad
  q_L(x) =   - \dfrac{1}{2^4 32 \pi^2}  \sum_{\mu \nu
\rho \sigma = \pm 1}^{\pm 4}\tilde\epsilon_{\mu \nu \rho \sigma}
\Tr\left[U_{\mu \nu}(x) U_{\rho \sigma}(x)\right],
\ee
where $S_W$ is the standard Wilson action, $U_{\mu \nu}$ the plaquette operator and $q_L$ a lattice version of the topological charge density. The latter needs renormalisation and the subscript on both $q_L$ and $\theta_L$ is a reminder that simulations are carried out in terms of unrenormalised parameters. Furthermore, $q_L$ is not topological and in particular not a total derivative. However, at this stage we simply take the action (\ref{eq:st}) as a given complex-valued lattice action, which we wish to study.

The action is complex for real $\theta_L$, while it is real for purely imaginary $\theta_L$ \cite{D'Elia:2012vv}. In the latter case we can therefore use both  Langevin dynamics (with gauge cooling, in order to keep the dynamics in SU(3), one could of course also use re-unitarisation) as well as algorithms based on importance sampling, such as hybrid Monte Carlo (HMC). 
Here we present {\em preliminary results} on a $6^4$ lattice, for $\theta_L^2 = 0, \pm 1, \pm 4$, i.e.\ for real and imaginary $\theta_L$. 
In order to gain confidence in our simulations, we show in Fig.\ \ref{fig:ExactResult} (right) the plaquette expectation value as a function of $\theta_L^2$. For $\theta_L^2\leq 0$, HMC results are indicated as well. We observe a smooth dependence on $\theta_L^2$ and agreement between the Langevin and HMC results when $\theta_L^2 \leq 0$, as it should be.

In Fig.\ \ref{fig:plaq} (left) we show the distributions of the real and imaginary part of the action at fixed real $\theta_L=2$, for three $\beta$ values. At larger $\beta$, gauge cooling is very effective in controlling the skirts of the distribution, at smaller $\beta$ less so. This problem, also present at nonzero baryon density  \cite{Seiler:2012wz}, requires a more careful analysis.
Finally, in Fig.\ \ref{fig:plaq} (right) we show the running average of the lattice topological charge, $Q_L=\sum_x q_L$ (without any topological cooling).
Note that for imaginary $\theta_L$, $\bra Q_L\ket$ is real, while for real $\theta_L$ it is imaginary: hence we show the real resp.\ imaginary part only.
For small $\theta$, the topological charge (density) is expected to satisfy
\be
\bra q_L\ket  = i\theta_L \chi_L + {\cal O}(\theta_L^3),
\ee
where $\chi_L$ is the lattice topological susceptibility. This $\theta_L$ dependence is  confirmed in Fig.\ \ref{fig:plaq} (right).
Interestingly, fluctuations for real/imaginary $\theta_L$ are very different, which remains to be understood.

\section{Summary}

Complex Langevin dynamics for gauge theories requires gauge cooling to control the exploration of the enlarged configuration space.  Efficient ways to  implement this employ adaptive cooling. We presented first results for SU(3) Yang-Mills theory in the presence of a $\theta$-term. While many questions still need to be addressed, this is, as far as we know,  the first time that simulations have been carried out directly for real $\theta$.

\acknowledgments

We thank Massimo d'Elia for providing the HMC code used in this work. 
Numerical results have been
obtained on computational resources provided by High Performance Computing (HPC) Wales.
This work is supported by STFC, the Royal Society, the Wolfson Foundation and the Leverhulme Trust.

\end{document}